\documentclass[twocolumn]{article} 
\usepackage[extreme]{savetrees}

\usepackage[utf8]{inputenc} 

\usepackage{geometry} 
\usepackage{graphicx} 

\usepackage{booktabs} 
\usepackage{array} 
\usepackage{paralist} 
\usepackage{verbatim} 
\usepackage{subfig} 

\usepackage{fancyhdr} 
\usepackage{sectsty}
\allsectionsfont{\sffamily\mdseries\upshape} 

\usepackage[nottoc,notlof,notlot]{tocbibind} 
\usepackage[titles,subfigure]{tocloft} 

\usepackage{amsmath}
\DeclareMathOperator{\sign}{sign}

\title{Fat Tails and Optimal Liability Driven Portfolios }
\author{Jan Rosenzweig\footnote{j.rosenzweig@pinetree-funds.com}}
\date{} 

\begin{document}
\maketitle

\section*{Abstract}
We look at optimal liability-driven portfolios in a family of fat-tailed and extremal risk measures, especially in the context of pension fund and insurance fixed cashflow liability profiles, but also those arising in derivatives books such as delta one books or options books in the presence of stochastic volatilities.\\
In the extremal limit, we recover a new tail risk measure,  {\it Extreme Deviation (XD)}, an extremal risk measure significantly more sensitive to extremal returns than CVaR. 
Resulting optimal portfolios optimize the return per unit of XD, with portfolio weights consisting of a liability hedging contribution, and a risk contribution seeking to generate positive risk-adjusted return.\\
The resulting allocations are analyzed qualitatively and quantitatively in a number of different limits.

{\bf Key words: ALM, LDI, fat tails, extremal risk measures}

\section{Introduction}

Portfolio optimization in the presence of fat tails has received a considerable amount of attention recently (see \cite{FTF,PLP} and references therein).

Briefly, it is known that, as the risk penalty moves from variance towards fat-tailed risk measures, the allocation becomes less dependent on the return of the component, and the resulting portfolio becomes more diversified \cite{FTF}. In the limit of extremal risk measures, the portfolio allocation becomes perfectly diversified, with the return dependence reducing to a simple in-out step function \cite{PLP}. 

In the context of LDI, and especially LDI for Solvency 2, there is also increased awareness of the importance of tail and extremal risk scenarios. While classical LDI considers portfolios that are optimal with respect to a variance-like risk measure such as the Value at Risk (VaR), there has been a considerable shift towards adopting tail based risk measures such as the Expected Shortfall or the Conditional Value at Risk (ES, CVaR) \cite{ldi}.

In this paper, we look to join the two approaches, by considering liability-driven portfolios with algebraic tail and extremal risk measures of \cite{FTF, PLP}. We do this in the classical LDI scenarios of pension funds and insurances, where the liabilities comprise of a series of fixed or almost fixed cash flows, but also in the context of derivatives books, where the liabilities comprise of derivatives written to clients, in the form of either delta one products or options.

The paper is organised as follows: Section 2 goes through the general analysis of the optimization problem and its solutions. Section 3 briefly justifies the extremal risk limit. Section 4 focuses on the classical pension fund and insurance LDI scenarios. Sections 5 deals with liabilities arising in client-facing derivative books, and Section 6 summarizes the conclusions.

\section{ALM portfolios}

The liability driven portfolio optimization problem is to find the optimal weights vector ${\bf w}=(w_{i})$ of the asset vector ${\bf A}=(A_{i})$, against a fixed liability portfolio $L$. Mathematically, assuming that the relevant moments exist and that they are finite, we are solving
\begin{equation}
{\bf w} . E \left(  d{\bf A} - {\bf r}_{A}dt \right) - dL + r_{L} dt - \lambda E \left[ {\bf w} . (d{\bf A} - {\bf r}_{A} dt) - dL + r_{L} dt \right]^{2k} \rightarrow \max \label{ldi} 
\end{equation}
for ${\bf w}$, where $E$ is the expectation operator, $d{\bf A}$ and $dL$ denote the respective returns of the asset vector and the liability, $\lambda > 0$ is a fixed risk tolerance, ${\bf r}_{A}$ and $r_{L}$ are the respective asset and liability funding rate vector and scalar, and $2k$ is a positive even integer. 

The expectation is taken ex-ante at time $t=0$, so that the current values of ${\bf A}$ and $L$ are known, their future returns $d{\bf A}$ and $dL$ are random, and we wish to influence the statistics of the returns distribution of the asset-liability portfolio. In abstract terms, the task is to maximise the returns of the joint portfolio, while minimizing its variability, for a range of measures of variability.

The choice of exponent as an even integer $2k$ restricts us to variance-like symmetric penalties, where large positive returns are penalised equally as large negative returns, as opposed to a skew-like penalty with a power of $2k-1$. The reason for this choice is twofold. First, any skew penalty term could become negative by switching from long to short holding. A negative skew term could not serve as an effective penalty, and the optimal portfolio would have weights that grow without bound.
Second, the standard negative skew in financial time series corresponds to frequent small positive returns, and infrequent large negative returns. While this is undesirable, the converse of this, which is frequent small negative returns, and infrequent large positive returns -  is generally not particularly desirable either. We therefore restrict ourselves to even exponents, while noting that it is nonetheless possible to extend the results to other exponents, as per  \cite{PLP}.

By letting $k>1$, we move away from  variance as a measure of portfolio risk, and this allowes us to capture fat-tailed \cite{FTF} and extremal risk measures \cite{PLP}. Our goal is to analyse the structure of the optimal portfolio as the risk measure changes from variance-like risk measures to fat-tailed and extremal risk measures.

Following \cite{FTF} and \cite{PLP}, we solve (\ref{ldi}) using an appropriate orthogonal decomposition of the asset vector, such as the ICA, kernel PCA or a neural network-based decomposition. 

Denoting the resulting orthogonal components $C_{1}, C_{2}, ...$ and their funding rates and expected  rates of return $r_{1}, r_{2},...$ and $\mu_{1}, \mu_{2},...$, we follow the  analysis from  \cite{PLP} to write the approximate formal solutions     for their respective weights $w_{1}, w_{2},...$. 

Briefly, in the direction of each $C_{i}$, (\ref{ldi}) is a polynomial in  $w_{i}$    of order $2k$, with up to $k$ local maxima. We are looking for its global maximum.

The problem simplifies considerably in two limits. The first limit, when the return term is small compared to the risk appetite, $(\mu_{i}-r_{i})/2k\lambda \ll 1$, at leading order gives a simple polynomial with a single maximum with multiplicity $2k$, reached at the weight $w_{i}$ that minimizes the $2k$-distance between $w_{i}dC_{i}$ and $dL$. This solution is known analytically \cite{lp}, and the global maximum is then found by its perturbation expansion in powers of $(\mu_{i}-r_{i})/2k\lambda$. Its first two terms are
\begin{equation}
w_{i}  = \rho_{i}\frac{\sigma_{i}}{\sigma_{L}} + \frac{1}{2 k (2k-1)\lambda} \frac{1}{E \left(d \hat{L}_{i}^{2k-2}     \right)} \frac{\mu_{i}-r_{i}}{\sigma_{i}^{2}}   +  O \left( \frac{\mu_{i}-r_{i}}{2k\lambda} \right)^{2},
\label{trivialweight}
\end{equation}
where $\sigma_{i}$   denotes the standard deviation of $C_{i}$,  $\rho_{i}$ its correlation with the liability,  $\sigma_{L}$ is the standard deviation of the liability, and $\hat{L}_{i}$ is the part of the liability that is orthogonal to $C_{i}$,
$$
\hat{L}_{i} = L - C_{i}  \rho_{i}\frac{\sigma_{i}}{\sigma_{L}} .
$$

The other limit of interest is  when the return term is large compared to the risk appetite, $(\mu_{i}-r_{i})/2k\lambda \gg 1$ . The analysis is then analogous to that described in \cite{PLP}; the global maximum is reached approximately where the leading term $w_{i}^{2k}$ balances the linear term  $w_{i}$, and the result is 
\begin{equation}
w_{i}  = \left[ \frac{(\mu_{i} - r_{i}) / 2 k \lambda  +  E \left[(dC_{j} - \mu_{j}dt) d\tilde{L}^{2k-1} \right]  }{E(dC_{j}-\mu_{j}dt)^{2k}}   \right] ^{1/(2k-1)} + O(1).
\label{weight}
\end{equation}
Where $d\tilde{L}=dL-r_{L} dt$ is the driftless part of $dL$, so that $E (d\tilde{L}) = 0$.

Note that (\ref{trivialweight}) and (\ref{weight}) only correspond to the same local maximum when $k=1$. Otherwise, they are different local maxima, whose status as a global maximum changes discontinuously as the risk tolerance parameter changes. 

The interpretation of the two maxima is straightforward; (\ref{trivialweight}) minimizes the risk generated by the liability, while (\ref{weight}) maximizes the return of the combined asset-liability portfolio. We therefore refer to (\ref{trivialweight}) as the {\it risk-avoiding} allocation, and to (\ref{weight}) as the {\it return-seeking} allocation.

Both solutions (\ref{trivialweight}) and (\ref{weight}) indicate two sources of allocation to a risky asset; a {\it return} term, allocated in accordance with its return and the risk appetite $\lambda$, and a {\it hedge}, a risk-tolerance-independent, return-independent term  depending on its mixed moments with the liability. 

No simplifying assumptions have been made on the nature of any of the processes, other than that the relevant expectations exist and are finite.

For the classical case $k=1$, the $O(1)$ error term in (\ref{weight}) vanishes, and (\ref{weight}) simplifies to the usual formula

\begin{equation}
w_{i} = \frac{1}{2\lambda } \frac{\mu_{i}-r_{i}}{\sigma_{i}^{2}} + \rho_{i} \frac{\sigma_{L}}{\sigma_{i}}
\label{classical}
\end{equation}
comprising of risky allocation in accordance to volatility-normalised Sharpe ratio, and the liability hedging allocation according to the hedge ratio. (\ref{classical})  is, of course, the same as (\ref{trivialweight}) for $k=1$, and the risk-avoiding and the return-seeking allocations are one and the same.

Qualitatively, the risk-avoiding allocation (\ref{trivialweight}) is independent of $k$. The allocation consists of the standard linear hedge $\rho_{i} \sigma_{i}/\sigma_{L}$  and a risk-adjusted term proportional to the volatility-normalised Sharpe ratio. The only dependence on $k$ comes in, at the leading two orders in the asymptotics, through the rescaling of the return-seeking term by a constant depending on $k$, which can be absorbed into rescaling the risk appetite at the level of a single component. We therefore omit further analysis of (\ref{trivialweight}) in this paper.

The goal of this paper is to analyze return-seeking allocations (\ref{weight}) for $k\ge 1$, and in the limit of $k \rightarrow \infty$.

\section{Extreme Deviation}

In this section, we elaborate on the choice of the penalty function, which is taken as the $2k$'th central moment of the returns distribution. As discussed above, this is a straightforward generalization of the variance, and the value of $k=1$ corresponds exactly to variance. As the value of $k$ increases, the penalty function is more sensitive to the tails of the distribuion.

An additional interpretation of the penalty function is found in the limit of $k\rightarrow \infty$. Say 
$$dx_{\infty} = \max_{i} |dx_{i}|< \infty.$$

Then
$$E \left( dx^{2k} \right) = \frac{1}{n} \sum_{j=0}^{n-1} dx_{i}^{2k}=  dx_{\infty}^{2k}  \frac{1}{n}  \sum_{j=0}^{n-1} \left( \frac{dx_{i}}{dx_{\infty} } \right)^{2k}$$
where all $|dx_{i}/dx_{\infty}| \le 1$.

\begin{figure}[h]
\centering
\includegraphics[width=7cm]{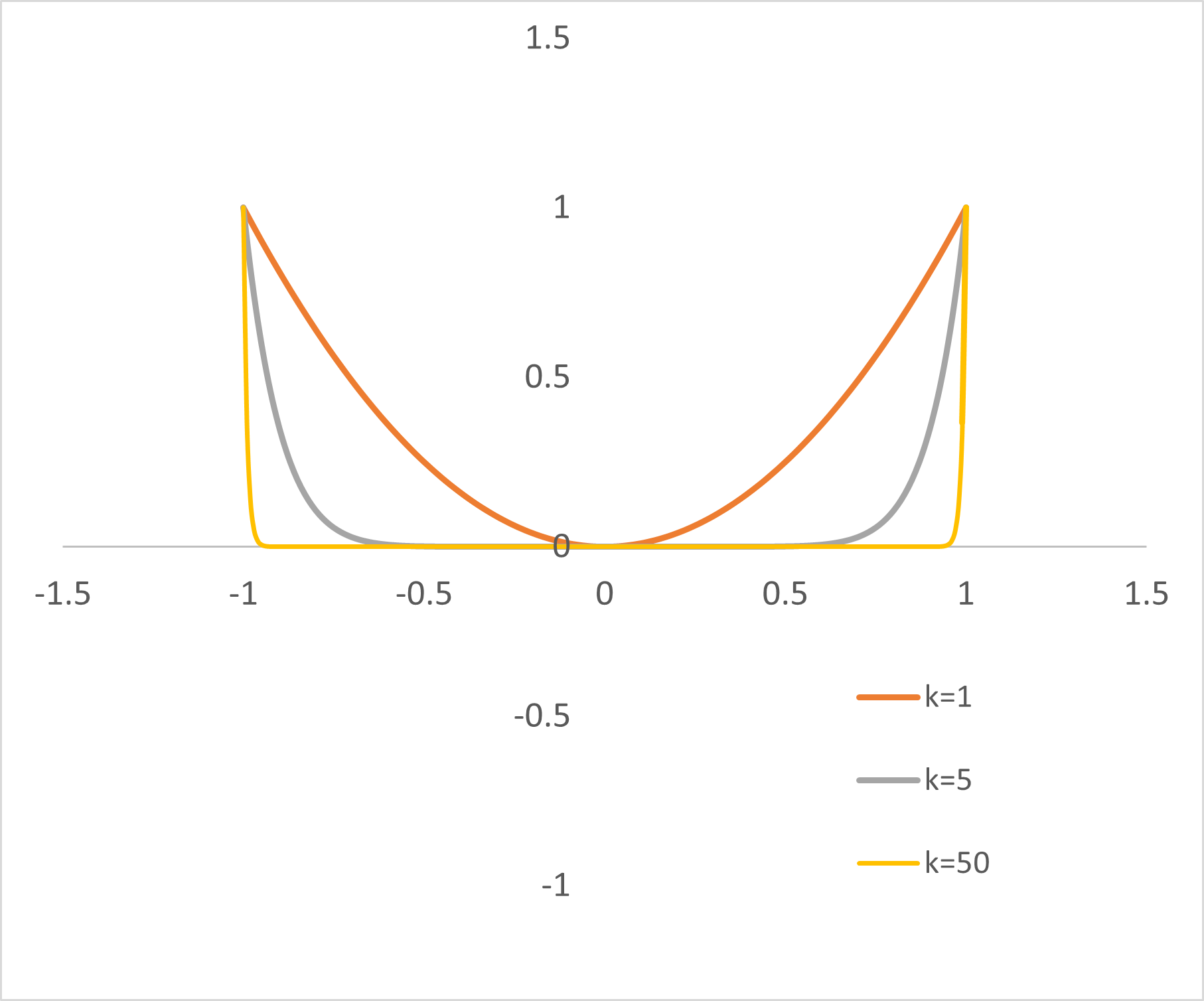}
\caption{Profiles of the power law damper}
\label{fig:hedged}
\end{figure}

See Figure \ref{fig:hedged} for a graph of the power-law function applied to each $dx_{i}/dx_{\infty}$. In essence, taking a small number to a high power works as a high-pass amplitude filter. Numbers close to $0$ vanish, and only numbers in a band around $\pm dx_{\infty}$ remain. The band becomes narrower as $k$ increases.

In the $k \rightarrow \infty$ limit,
$$\left( E dx^{2k} \right)^{1/2k}=  dx_{\infty}  \left( \frac{1}{n} \sum_{j=0}^{n-1} \left( \frac{dx_{i}}{dx_{\infty} } \right)^{2k} \right)^{1/2k}\approx$$
$$\approx dx_{\infty}  \left(\frac{1}{n} \sum_{|dx_{j}|=dx_{\infty}} 1^{2k} \right)^{1/2k} = dx_{\infty} \left( \frac{n_{0}}{n} \right)^{1/2k} \rightarrow  dx_{\infty}$$
where $n_{0}$ is the nuber of points on which the maximum norm is reached. We therefore get a straightforward interpretation of our high order moments as first order estimates of extreme de-meaned returns.

The limit $k\rightarrow \infty$ is in itself a risk measure, belonging to the family of deviation risk measures \cite{deviation}, which we call {\it Extreme Deviation (XD)}. It is the expected maximum absolute return of the de-meaned returns distribution, i.e. maximum absolute deviation around the mean.

\begin{figure}[h]
\centering
\includegraphics[width=9cm]{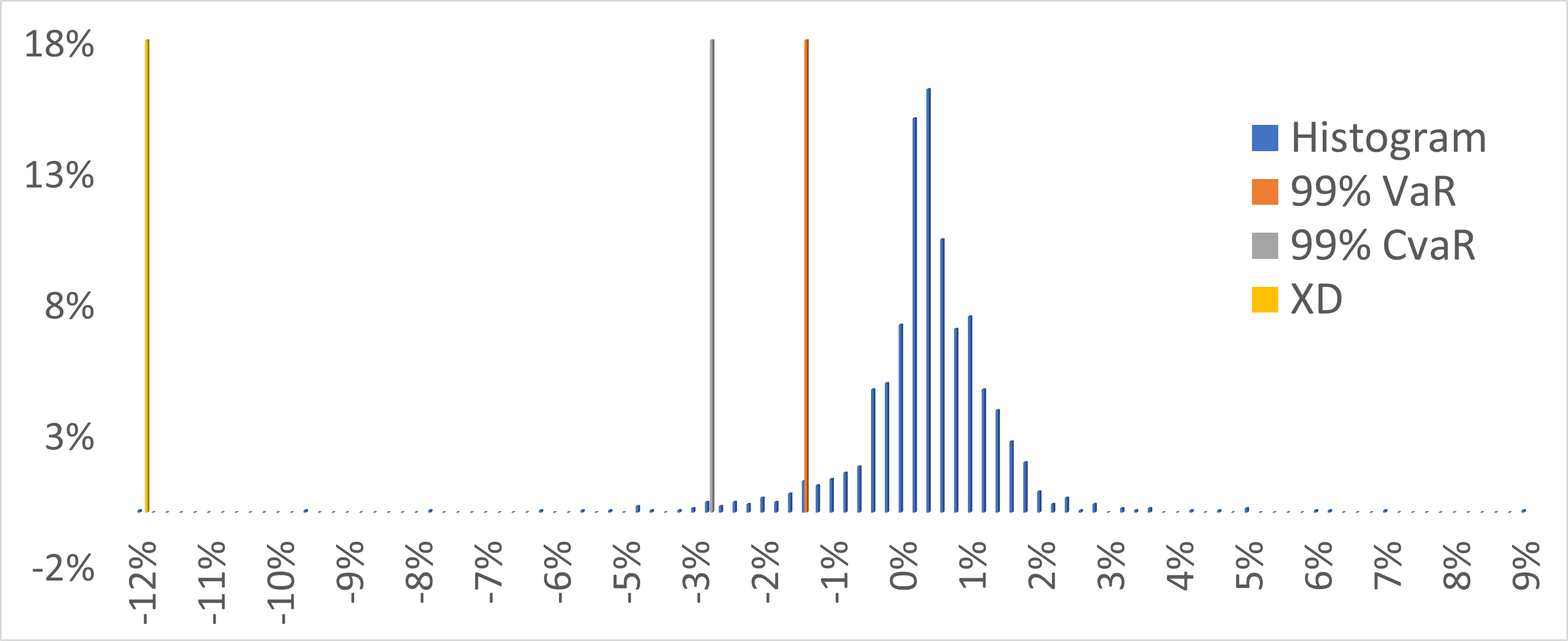}
\caption{A comparison of 99\% VaR, 99\% CVaR and XD for daily SPX returns 2017-2021.}
\label{fig:riskmeasures}
\end{figure}

A comparison between VaR, CVaR and XD on the daily returns of the SPX from 2017-2021 is shown in Figure \ref{fig:riskmeasures}. 

As a risk measure, it is fairly straightforward to establish that XD is translation invariant, positive homogenous, sublinear and positive; 
$$
XD ( \alpha + \beta Z) = \max_{i} \left| d (\alpha +  \beta\   z_{i} ) \right| = |\beta|\ dz_{\infty} = |\beta|\ XD(Z)
$$
$$
XD(Y+Z) = \max_{i} | dy_{i} + dz_{i} | \le \max_{i} \left(| dy_{i}| + | dz_{i} | \right) \le $$ $$ \le \max_{i} | dy_{i}| + \max_{j} | dz_{j} | = XD(Y)+XD(Z).
$$
$$ XD(Z) = 0 \implies \max_{i} |dz_{i} | = 0  \implies dz_{i} = 0\ \forall i \implies Z \equiv const.$$

In summary, XD is an unconditional extremal risk measure such that, for any random variable $Z$,
$$
VaR_{p}(Z) \le CVaR_{p}(Z) \le XD(Z)
$$
for any probability $p$. Its importance in our context is that, (i) it can be estimated robustly from the high order moments, and (ii) it provides the interpretation of the optimal portfolio in the limit of $k \rightarrow \infty$ as the portfolio with maximum return per unit of XD.

\section{From Least Squares to Minimax}

There is a qualitative change of behaviour in the character of the optimization problem (\ref{ldi}) as the value of the exponent $k$ increases from $k=1$ towards $k=\infty$. Specifically, for $k=1$, the optimization is a least squares problem; for $k>1$ it is $2k$-norm optimization; and, for $k=\infty$, it is the $\infty$-norm, or minimax optimization.

\begin{figure}[h]
\centering
\includegraphics[width=9cm]{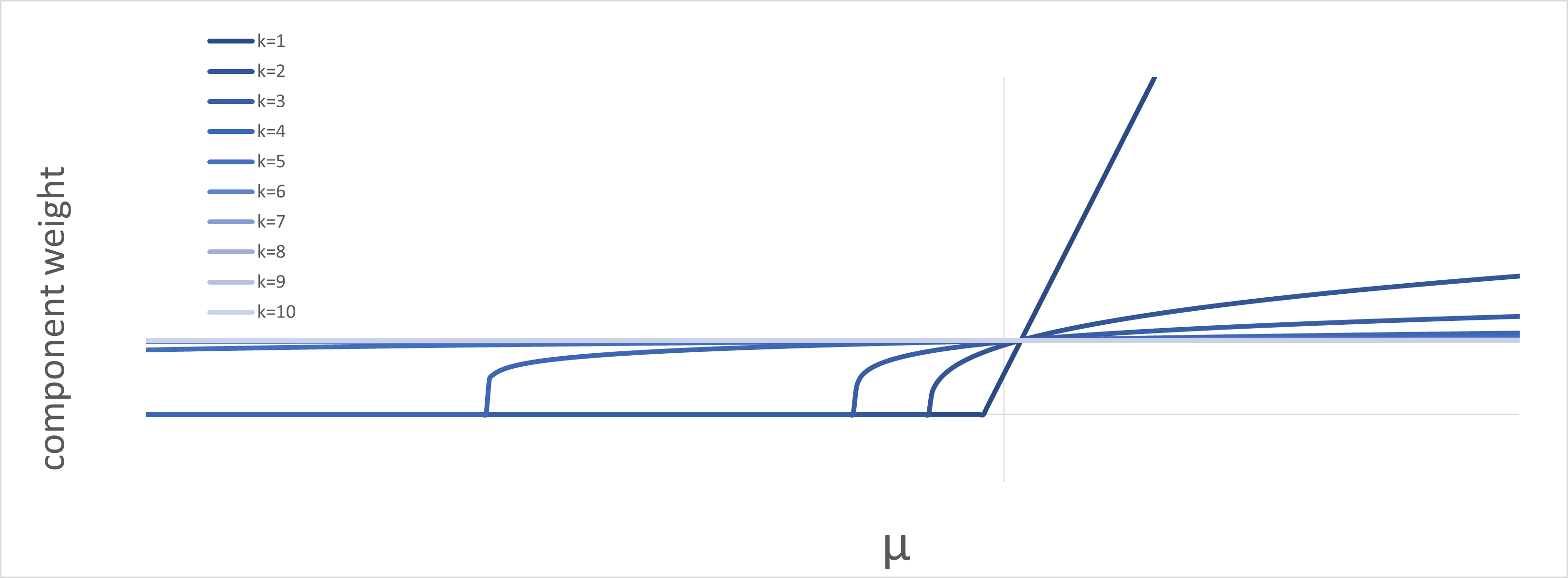}
\caption{Graph of  (\ref{weight}); component weight $w$ as function of its return $\mu$ for varying $k$.}
\label{fig:weights}
\end{figure}

To get some sense for how the optimal allocation depends on the risk appetite, we can look at (\ref{weight}) a bit more closely as a function of $\mu$. The graph of (\ref{weight}) for varying $\mu$ and $k$ is shown in Figure \ref{fig:weights}.  By inspection, we can see that the crossed moment term effectively generates a correction to the hurdle rate $r$, given as 
\begin{equation}
r_{c} = 2 k \lambda  E \left[(dC_{j} - \mu_{j}dt) d\tilde{L}^{2k-1} \right] 
\label{r0}
\end{equation}
and (\ref{weight}) can then be re-written as
\begin{equation}
w = \left[ \frac{1}{2 k \lambda} \frac{\mu - (r - r_{c})}{ E(dC-\mu dt)^{2k}} \right]^{1/(2k-1)}
\label{govvieweightt}
\end{equation}

In other words, the weight profile in terms of the return is still, as shown in  in Figure \ref{fig:weights}, the familiar pattern from \cite{PLP} of a hockey stick collapsing to a step function as $k$ progresses from $1$ to $\infty$, but, this time, with a hurdle rate that decreases exponentially as $k$ increases.

\section{LDI and Delta one products}

For Delta one products, the calculation of the ratio of the moments $ E \left[(dC_{j} - \mu_{j} dt) dL^{2k-1} \right] $ and $E(dC_{j}-\mu_{j}dt)^{2k}$   is straightforward if the underlying processes are Gaussian \cite{moments}; using
\begin{equation}
\begin{array}{cl}
dC_{i} & = \mu_{i} dt + \sigma_{j} dW_{i}\\
dL & = \sigma_{L} dW \\
dW_{i} dW &= \rho_{i} dt
\end{array}
\label{gaussianpowers}
\end{equation}
and orthogonalizing 
\begin{equation}
\begin{array}{cl}
dW_{i} &= \rho_{i} dW  + \sqrt{1-\rho_{i}^{2}} dZ_{i}\\
 dZ_{i} dW_{i} & = 0
\end{array}
\label{orho1}
\end{equation}
we get 
\begin{equation}
(dC_{i} - \mu_{i}dt) dL^{2k-1} = \sigma_{i} \sigma_{L}^{2k-1} \rho_{i}\ (2k-1)!!\ dt^{k} 
\label{summation}
\end{equation}

Since
$$ E(dC_{i}-\mu_{i})^{2k} =   (2k-1)!!\  \sigma_{i}^{2k} dt^{k},$$
\begin{equation}
\frac{ E \left[(dC_{i} - \mu_{i}dt) dL^{2k-1} \right]}{E(dC_{i}-\mu_{i})^{2k}}  =  \rho_{i} \left( \frac{\sigma_{L}}{\sigma_{i}} \right)^{2k-1}
\label{deltaoneratio}
\end{equation}

We can substitute (\ref{deltaoneratio}) back into (\ref{weight}) to get the full allocation weight as
\begin{equation}
w_{i} = \left[ \frac{1}{2 k \lambda} \frac{\mu_{i} - r_{i}}{ (2k-1)!! \sigma_{i}^{2k}} + \rho_{i} \left( \frac{\sigma_{L}}{\sigma_{i}} \right)^{2k-1}  \right]^{1/(2k-1)}
\label{deltaoneweight}
\end{equation}

\begin{figure}[h]
\centering
\includegraphics[width=9cm]{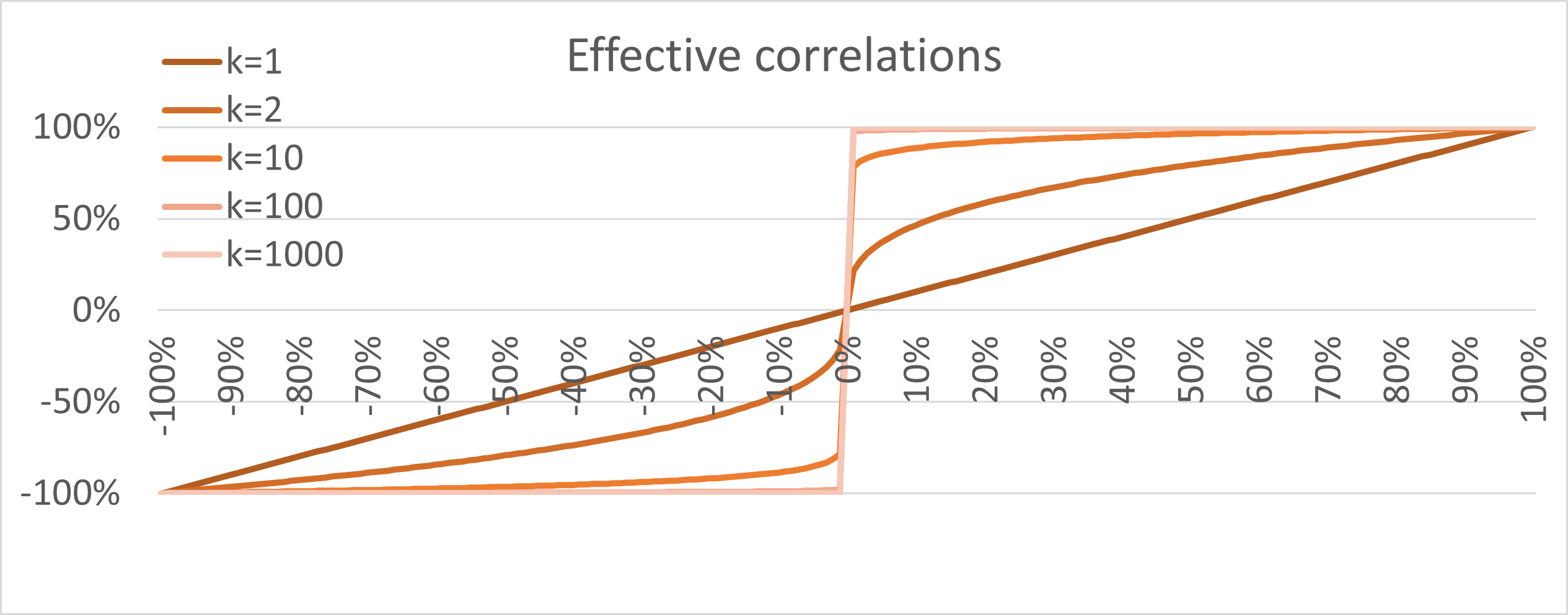}
\caption{Effective correlations for increasing vaues of $k$.}
\label{fig:correl}
\end{figure}

For large $k$ and fixed $\lambda$, the hedge ratio simplifies to
\begin{equation}
 w_{i} = \rho_{i}^{1/(2k-1)} \frac{\sigma_{L}}{\sigma_{i}} 
\label{hedgeratio2k}
\end{equation}
in other words, this limit has an "effective correlation" used for hedging, equal to the actual correlation raised to the power-law power of $1/(2k-1)$. The profile of this effective correlation is shown in Figure \ref{fig:correl}.

The the effective correlation term in the hedging ratio (\ref{hedgeratio2k}) has a simple limiting value for the weight $w_{i}$ as $k \rightarrow\infty$ with $\lambda$ fixed of
\begin{equation}
w_{i} \approx \sign (\rho_{i}) \frac{\sigma}{\sigma_{i}},
\label{hedgelimit}
\end{equation}
so all positive correlations are effectively treated as constant and equal to $1$, all negative correlations likewise as constant and equal to $-1$, and correlation 0 remains 0. 
This is somewhat in line with long-standing trading practice \cite{giller}. In practice, one would select a finite-width band around correlation 0 where the effective correlation would be deemed to small to count, and it would be set to $\pm 1$ outside this band.

We note that the limits (\ref{hedgeratio2k}) and (\ref{hedgelimit})  break the conditions under which (\ref{weight}) is the global maximum of (\ref{ldi}), and that this simple effective correlation therefore never actually appears as a hedging correlation. It is therefore useful as an intuitive shorthand, but it should not be used as an actual formula to construct portfolios.

The correct limit should ensure that (\ref{weight}) remains the global maximum; it is taken by using  (\ref{govvieweightt}) with the effective hurdle rate 
\begin{equation}
r_{c} = 2 k (2k-1)!!\  \lambda \rho_{i} \left( \frac{\sigma_{L}}{\sigma_{i}} \right)^{2k-1}  .
\label{r0doption}
\end{equation}
and it  results in a flat positive allocation with no hurdle rate.

\section{Options}

We now turn to the case where the liability is an option on the hedging underlyings $S_{1}, S_{2},...$, with its value denoted $\Omega(S_{1}, S_{2},...)$.

Standard Ito's lemma gives the dynamics of $\Omega$ as
\begin{equation}
d\Omega = \left[ \frac{\partial \Omega}{\partial t} + \frac{1}{2} \sigma_{i}^{2} \frac{\partial^{2} O}{\partial S_{i}^{2} } \right] dt + \sigma_{i} \frac{\partial \Omega}{\partial S_{i}} dW_{i} + \alpha_{i} \frac{\partial \Omega}{\partial \sigma_{i}} dZ_{i}
\label{ito}
\end{equation}
where we now introduce additional terms for $\alpha_{i}$, the vol-of-vol of $S_{i}$, and the stochastic vol factor $dZ_{i}$, and the summation convention applies. The smiles come from the vol-of-vol parameters $\alpha_{i}$, and the skews come from the spot-vol correlations,
\begin{equation}
dW_{i}dZ_{j} = \delta_{ij} q_{i} dt.
\label{spotvol}
\end{equation}

The randomness of the option process comes from the randomness of the spot processes, mediated by the respective deltas, and the randomness of the vol processes, mediated by the respective vegas.

Orthogonalizing the variance factor as
\begin{equation}
\begin{array}{cl}
dZ_{i} &= q_{i} dW_{i}  + \sqrt{1-q_{i}^{2}} dY_{i}\\
 dZ_{i} dY_{i} & = 0,
\end{array}
\label{qrho1}
\end{equation}

we go through the same motions as in (\ref{summation}); based on (\ref{ito}), the correlation between the option process $d\Omega$ and he component process $dC_{i}$ is 
$$
\rho_{i} = \frac{\partial \Omega}{\partial S_{i}}  + q_{i}\frac{\alpha_{i}}{\sigma_{i}} \frac{\partial \Omega}{\partial \sigma_{i}},
$$
which we can plug into (\ref{deltaoneweight}) to get the  moment ratio term as
\begin{equation}
\frac{ E \left[(dC_{j} - \mu_{j}dt) d\Omega^{2k-1} \right]}{E(dC_{j}-\mu_{j})^{2k}} =   \frac{\partial \Omega}{\partial S_{i}}  + q_{i}\frac{\alpha_{i}}{\sigma_{i}} \frac{\partial \Omega}{\partial \sigma_{i}}      
\label{optionratio}
\end{equation}

In other words, the moment ratio has a component equal to the skew-adjusted delta. The skew adjustment is the usual product of spot-vol correlation, vol-of-vol to vol ratio, and the option vega. Plugging this back into  (\ref{weight}), we get the full weight as 
\begin{equation}
w_{i} = \left[ \frac{1}{2 k \lambda} \frac{\mu_{i} - r_{i}}{ (2k-1)!! \sigma_{i}^{2k}} +  \frac{\partial \Omega}{\partial S_{i}}  + q_{i}\frac{\alpha_{i}}{\sigma_{i}} \frac{\partial \Omega}{\partial \sigma_{i}}     \right]^{1/(2k-1)}
\label{optionweight}
\end{equation}

The classical case of $k=1$ reduces to 
\begin{equation}
w_{i} = \frac{1}{2\lambda } \frac{\mu_{i}-r_{i}}{\sigma_{i}^{2}} + \frac{\partial \Omega}{\partial S_{i}} +  q_{i} \frac{\alpha_{i}}{\sigma_{i}}  \frac{\partial \Omega}{\partial \sigma_{i}},
\label{optionk2}
\end{equation}
and the (incorrect) limit of large $k$, fixed $\lambda$  is analogous to (\ref{hedgeratio2k}), as
\begin{equation}
w_{i} \approx \left[ \  \frac{\partial \Omega}{\partial S_{i}}  + q_{i}\frac{\alpha_{i}}{\sigma_{i}} \frac{\partial \Omega}{\partial \sigma_{i}}     \right]^{1/(2k-1)}
\label{hedgeratio2koption}
\end{equation}

The effective hurdle rate for  (\ref{govvieweightt}) is
\begin{equation}
r_{c} = 2 k (2k-1)!!\  \lambda \left( \frac{\partial \Omega}{\partial S_{i}} +  q_{i} \frac{\alpha_{i}}{\sigma_{i}}  \frac{\partial \Omega}{\partial \sigma_{i}} \right) .
\label{r0doption}
\end{equation}

\section{Examples}

\subsection{Pension fund LDI}

\begin{figure}[h]
\centering
\includegraphics[width=9cm]{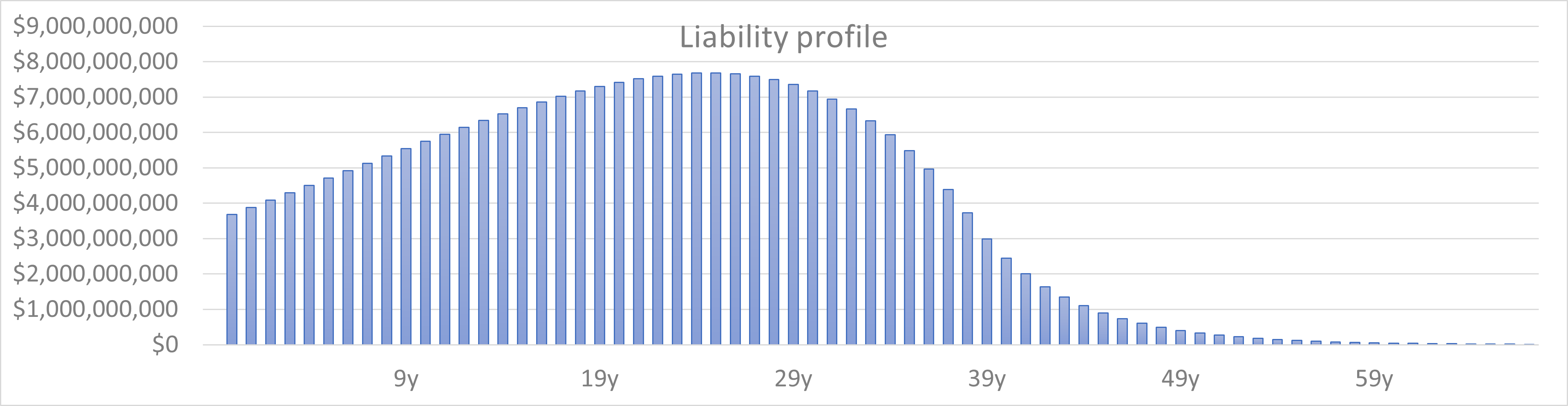}
\caption{A pension fund liability profile.}
\label{fig:profile}
\end{figure}

We use an example of an actual liability profile of a Middle-Eastern pension fund, depicted in Figure \ref{fig:profile}.

\begin{figure}[h]
\centering
\includegraphics[width=9cm]{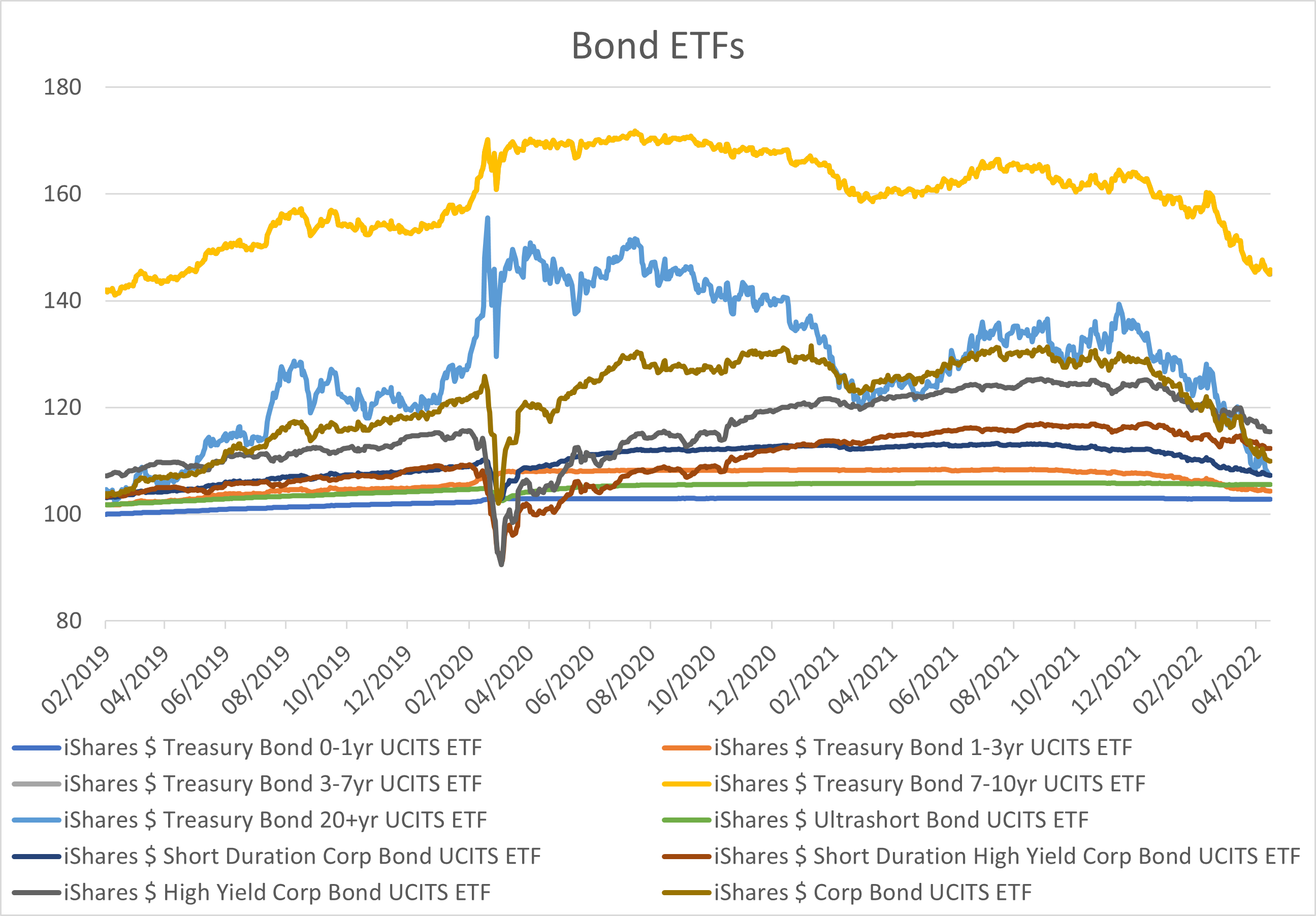}
\includegraphics[width=8cm]{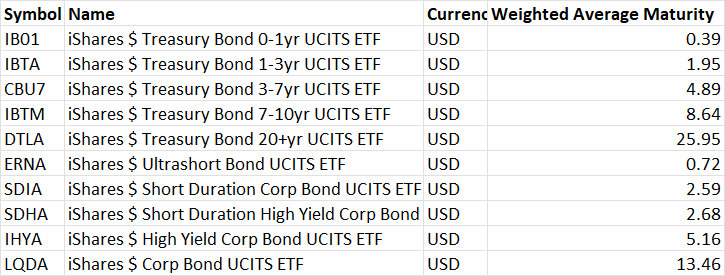}
\caption{Bond ETFs used for hedging the liability profile from Figure \ref{fig:profile}.}
\label{fig:ETFtable}
\end{figure}

The hedging universe consists of ten iShares bond ETFs, shown in Figure \ref{fig:ETFtable}. The ETF universe was chosen to cover the USD treasury curve, USD corporate bonds and USD high yield bonds. The time period under observation was 3 years, from 20th February 2019 until 19th February 2022. This time period includes the time of significant shock in the bond markets in March 2020 due to the Covid-19 pandemic, which serves as an example of a fat-tailed event.

We applied ICA decomposition to the return vectors of these ten bond ETFs to generate the independent components. The ICA package used was \texttt{fastICA} as implemented in the Python library \texttt{scikit-learn}.

\begin{figure}[h]
\centering
\includegraphics[width=9cm]{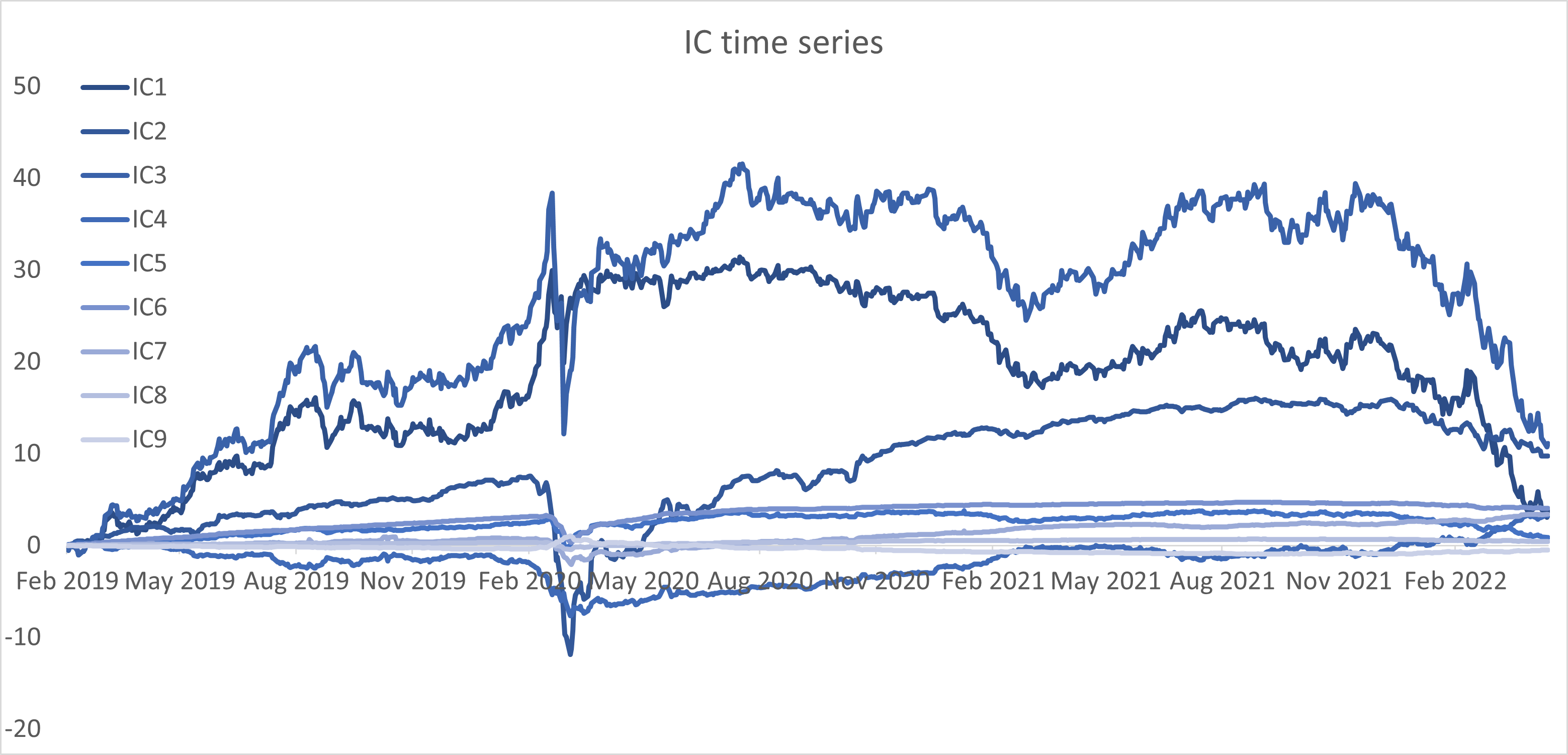}
\includegraphics[width=9cm]{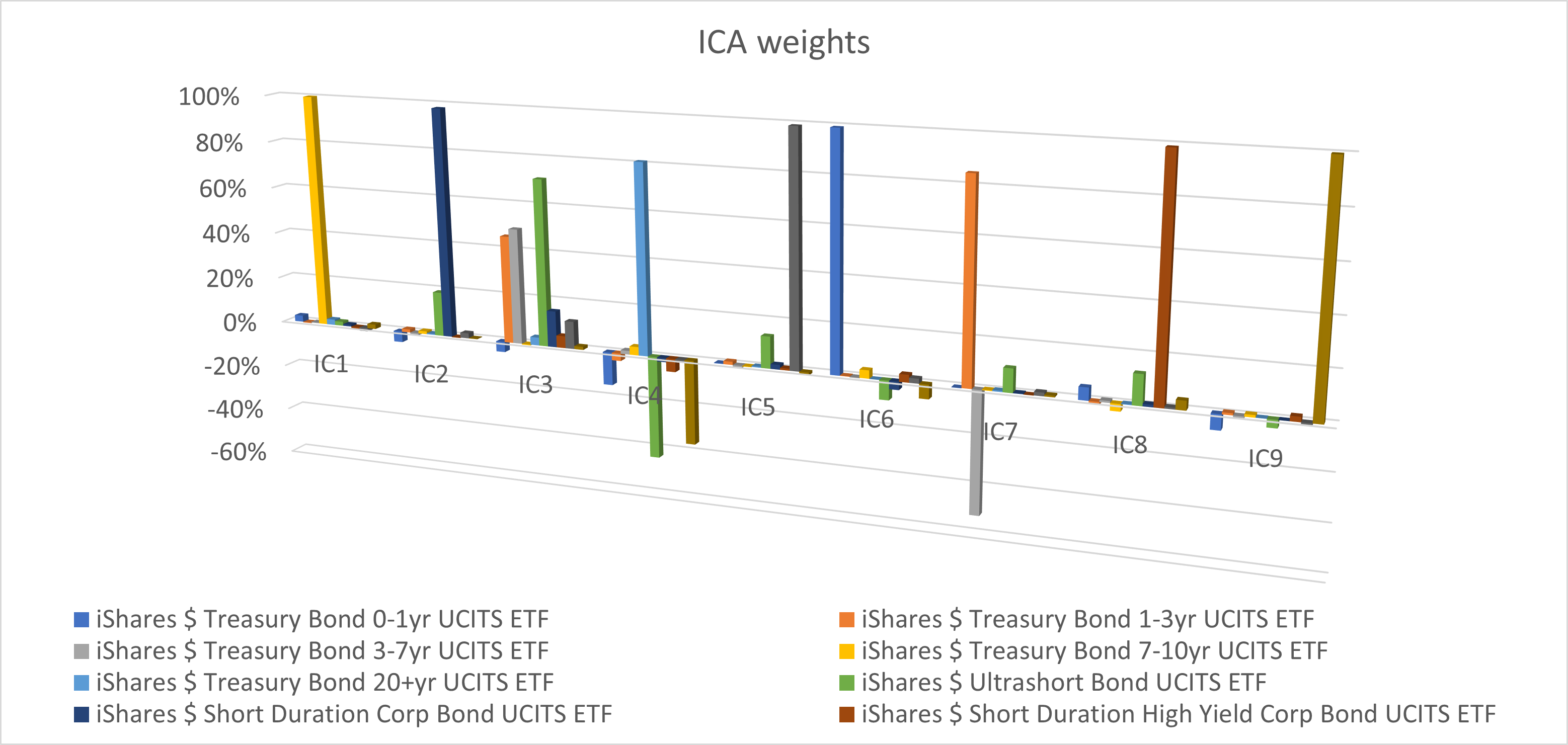}
\includegraphics[width=8cm]{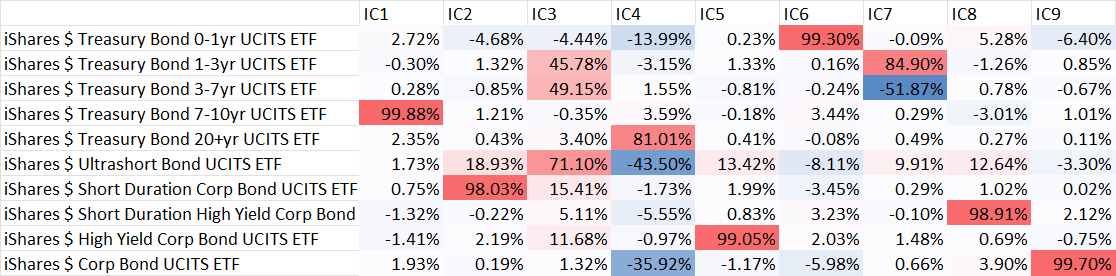}
\caption{Independent Component decomposition of the bond ETF universe.}
\label{fig:ICtable}
\end{figure}

Even though there are ten ETFs in the universe, the rank of their return vectors is only 9, indicating that one is redundant. Consequently, ICA returned nine independent factors. Their weights and trajectories are shown in Figure \ref{fig:ICtable}.

As seen from Figure \ref{fig:ICtable}, the ICs all have a straightforward interpretation, namely: IC6 describes the short end of the treasury curve; IC3 describes the parallel shift of the 1y7y treasury curve; IC7 describes the 1y10y steepener of the treasury curve; IC1 describes the 10y part of the treasury curve; IC4 describes the 20y part of the treasury curve; IC2 describes the short end of the corporate curve; IC9 describes the long end of the corporate curve; IC8 describes the short end of the high yield curve; and IC5 describes the long end of the high yield curve.

The NPV of the liability profile was discounted using treasury yields.

\begin{figure}[h]
\centering
\includegraphics[width=9cm]{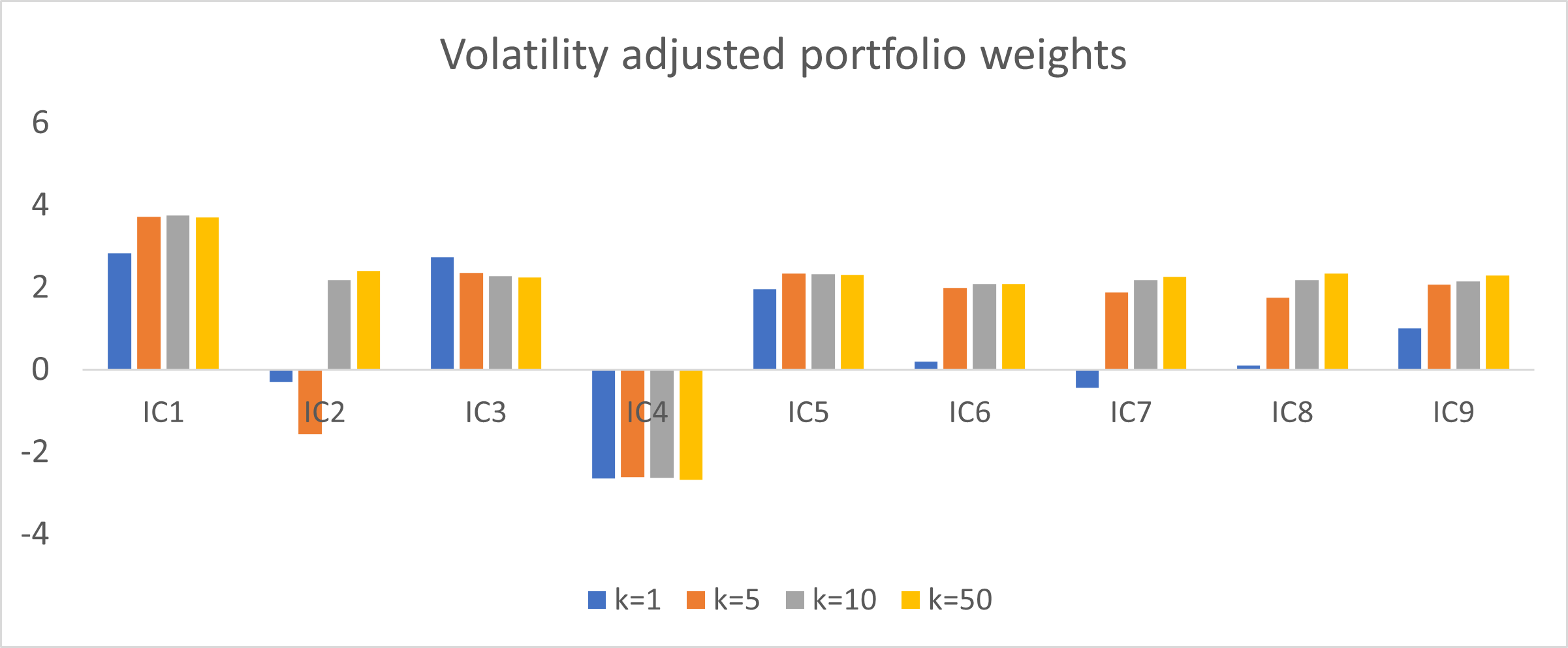}
\includegraphics[width=9cm]{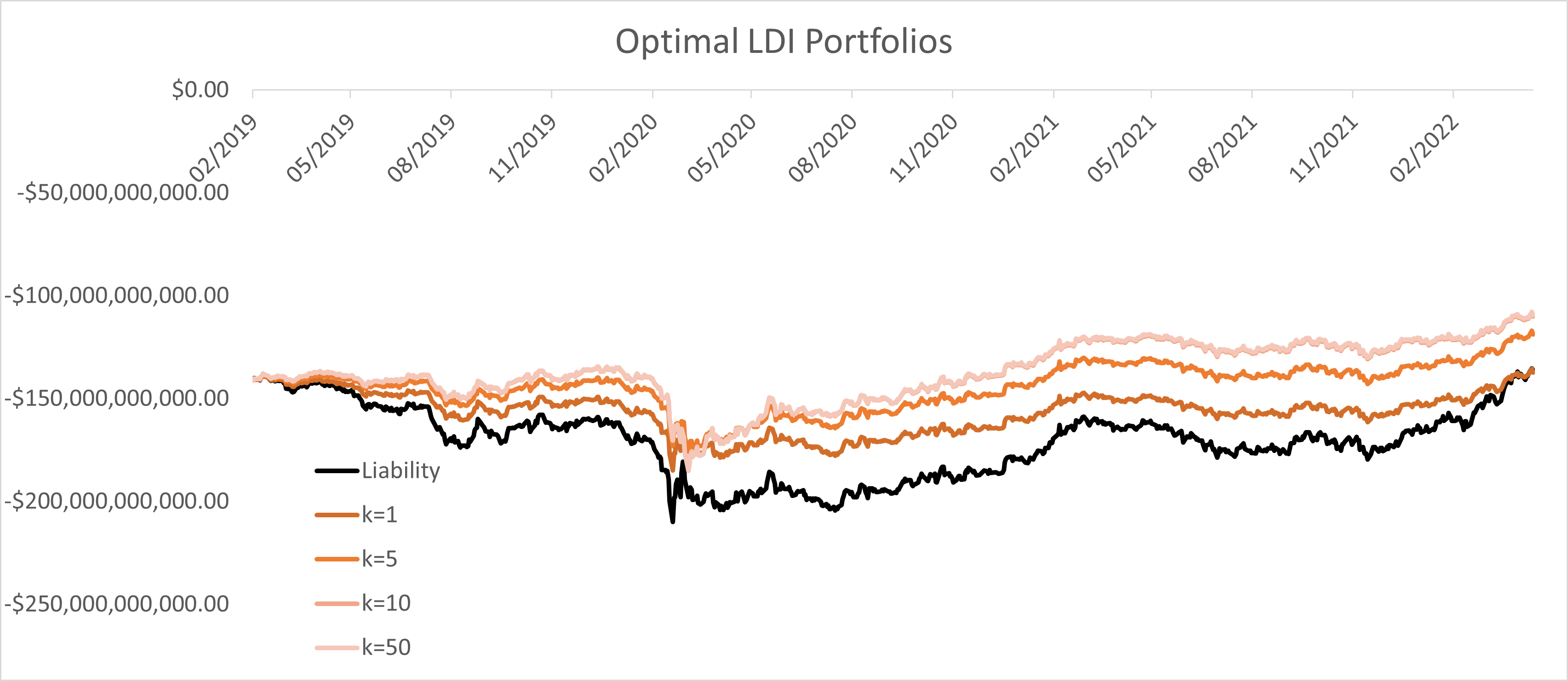}
\caption{Optimal LDI portfolios for increasing $k$, covering the Covid-19 induced shocks in 2020. Portfolio weights in terms of the ICs (top), and portfolio trajectories (bottom).}
\label{fig:LDItable}
\end{figure}

We composed optimal portfolios according to (\ref{weight}) for values of $k=1,5,10,50$ and the risk appetite $\lambda = 1\%$. The trajectories of the resulting portfolios are shown in Figure \ref{fig:LDItable}.

As seen from Figure \ref{fig:LDItable}, all choices of $k$ significantly lower the volatility of the liability portfolio, and this effect improves with increasing $k$. This is almost entirely due to high $k$ absorbing the Covid-19 related shocks in March 2020. The trajectories for increasing $k$ converge rapidly, and the curves for $k=10$ and $k=50$ are next to indistinguishable from each other.

A closer look at the weights distribution in Figure \ref{fig:LDItable} reveals that the most significant effect of increasing $k$ is adding allocations to corporate and high yield factors contained in IC2, IC6, IC7 and IC8. The liability is discounted at treasury yields, so corporate and high yield factors contribute diversification, rather than hedging. 

There is a small curiosity in the fact that one high yield factor, IC5, does find its way into the variance minimizing $k=1$ portfolio, instead of, say, the comparable treasury factor in IC7. The explanation for this is somewhat technical; namely, the 1y7y  steepener in IC7 does not contribute to hedging due to the straight line shape of the liability profile over the entire 1y20y bucket, and it is only added to high $k$ portfolios as a diversifier. The appearance of  the 5y high yield factor in IC5 is an unrelated artefact of the selected time period of observation.

\subsection{Option hedging}

We looked at the eamples of  exchange traded vanilla options on five NYSE underlyings: AAPL, MSFT, GE, GM and XOM. All options were written on the 28th February 2022, and expired on the 8th April 2022. The option position considered was a short ATM straddle based on  Close prices on the 25th February, which produced the strikes 170 for AAPL, 300 for MSFT, 95 for GE, 47 for GM, and 78 for XOM. The tickers for the options used are \texttt{AAPL220408C00170000},	\texttt{AAPL220408P00170000}, \texttt{MSFT220408C00300000},	\texttt{MSFT220408P00300000},\texttt{GE220408C00095000},	\texttt{GE220408P00095000}, \texttt{GM220408C00047000},	\texttt{GM220408P00047000},  \texttt{XOM220408C00078000}, and	\texttt{XOM220408P00078000}.

We compared the PnL of daily delta hedging to those from a hedging strategy with weights given by (\ref{optionweight}).

For delta hedging, option deltas were calculated based on daily  Close, and delta hedging was also performed at the same daily Close. A fixed trading cost of $0.06\%$ was applied to all transactions. While this assumes some forward knowledge on behalf of the book runner, in being able to trade at the same price that the delta was derived from, it is a reasonable assumption in practice. In practice, the deltas would be calculated from market prices prior to the Close, and, for liquid stocks, they would not significantly differ from the deltas obtained from Close. The delta used for the delta hedging was simple Black-Scholes delta at option strike, with the implied volatility taken as the average of the call and the put volatility derived from the option close prices. 

For the risk-seeking component in  (\ref{optionweight}), we used the mean return and moments of the daily returns of the underlying from the two months prior to the option being written, namely 3rd January 2022 until 25th February 2022. The risk appetite parameter was taken as $\lambda = 1\%$. The exponents were $k=1,5,10,50,100$. The expected return based on the 3rd January to 25th February bucket was negative for AAPL, MSFT and GM, and positive for GE and XOM. Realised return in the 28th February to 8th April bucket was negative for GE and GM, and positive for AAPL, MSFT and XOM.

\begin{figure}[h]
\centering
\includegraphics[width=6cm]{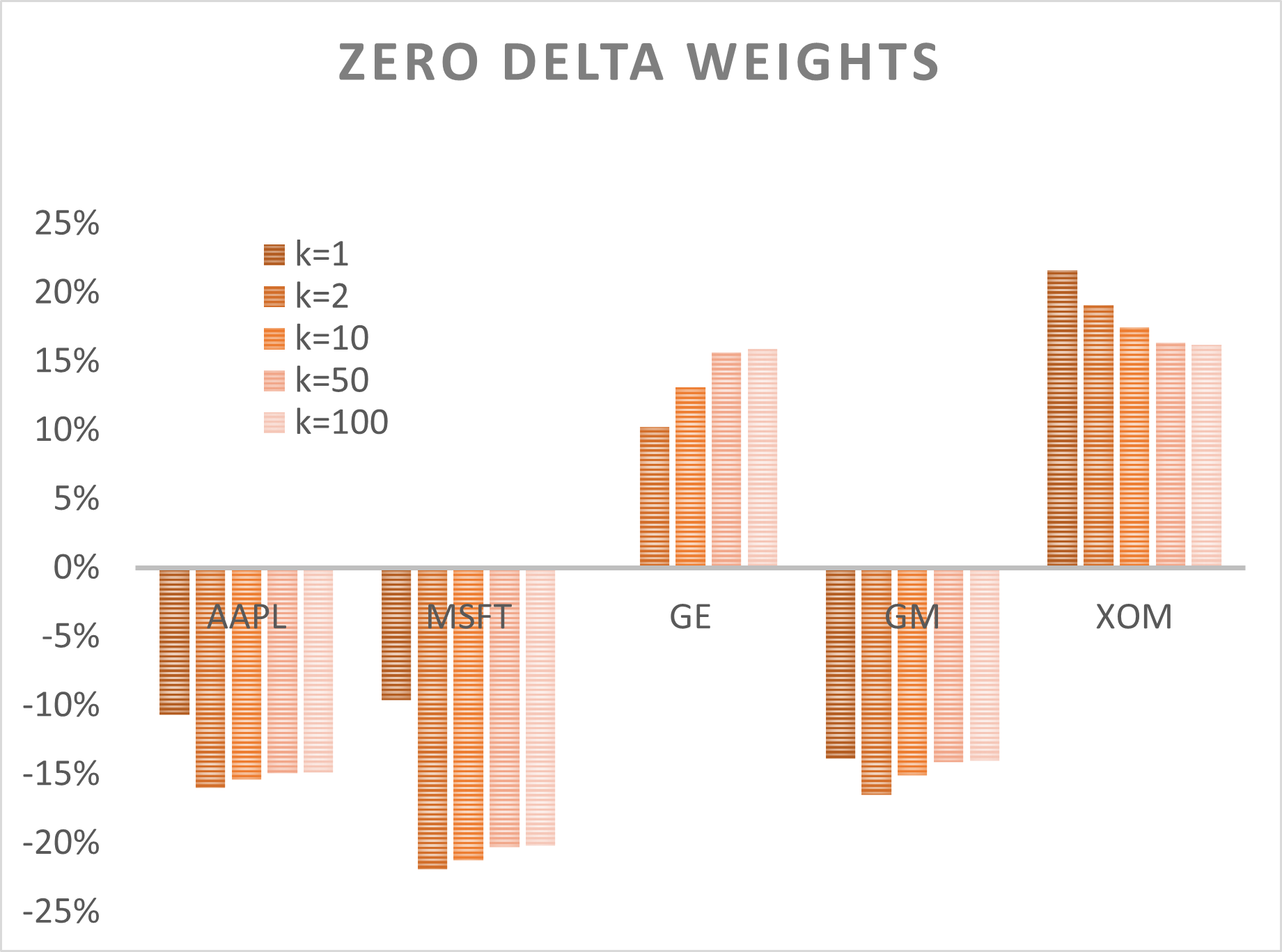}
\caption{Zero delta option  hedging weights for $k=1,2,10, 50, 100$, for AAPL, MSFT, GE, GM and XOM.}
\label{fig:optionweights}
\end{figure}

\begin{figure}[h]
\centering
\includegraphics[width=4cm]{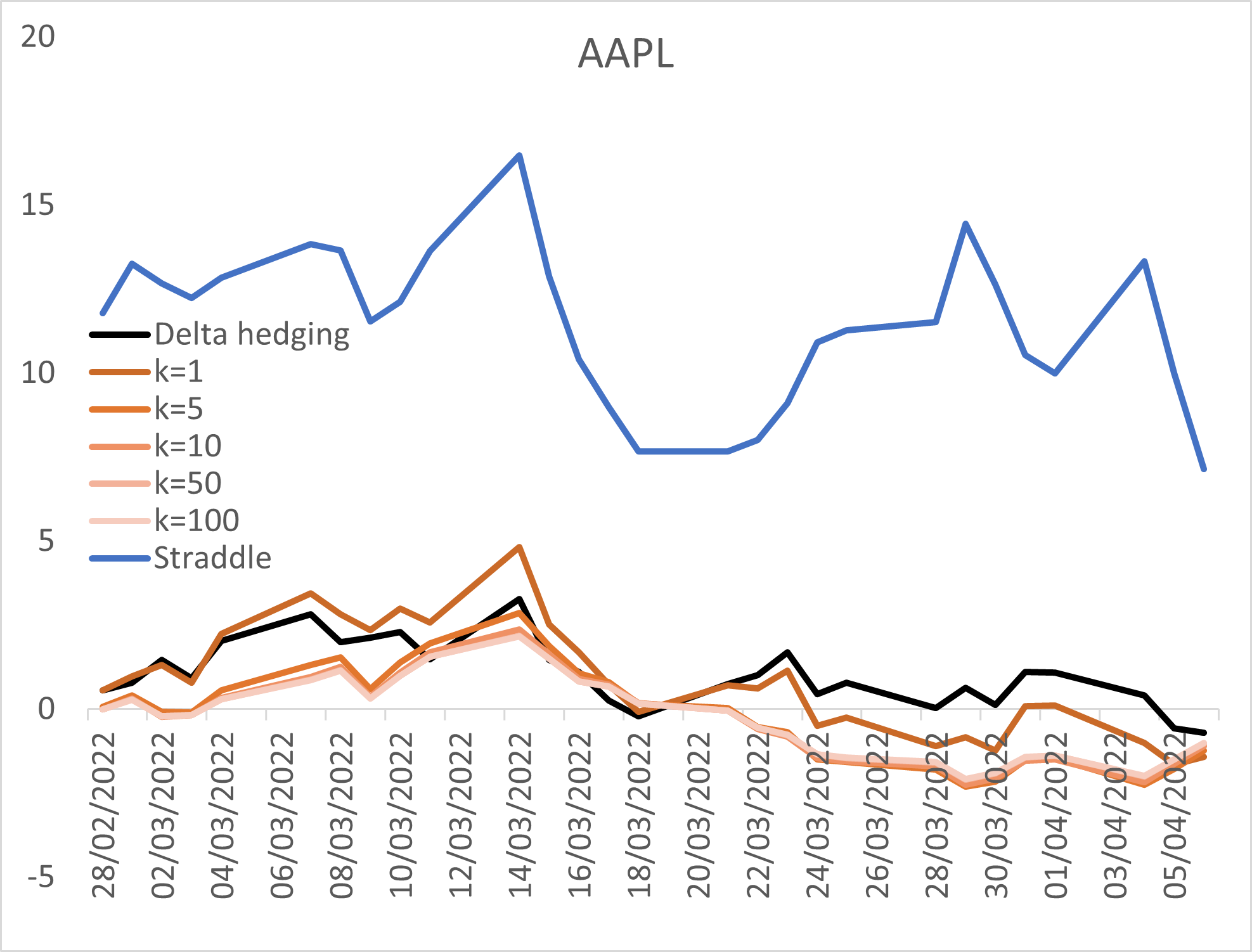}
\includegraphics[width=4cm]{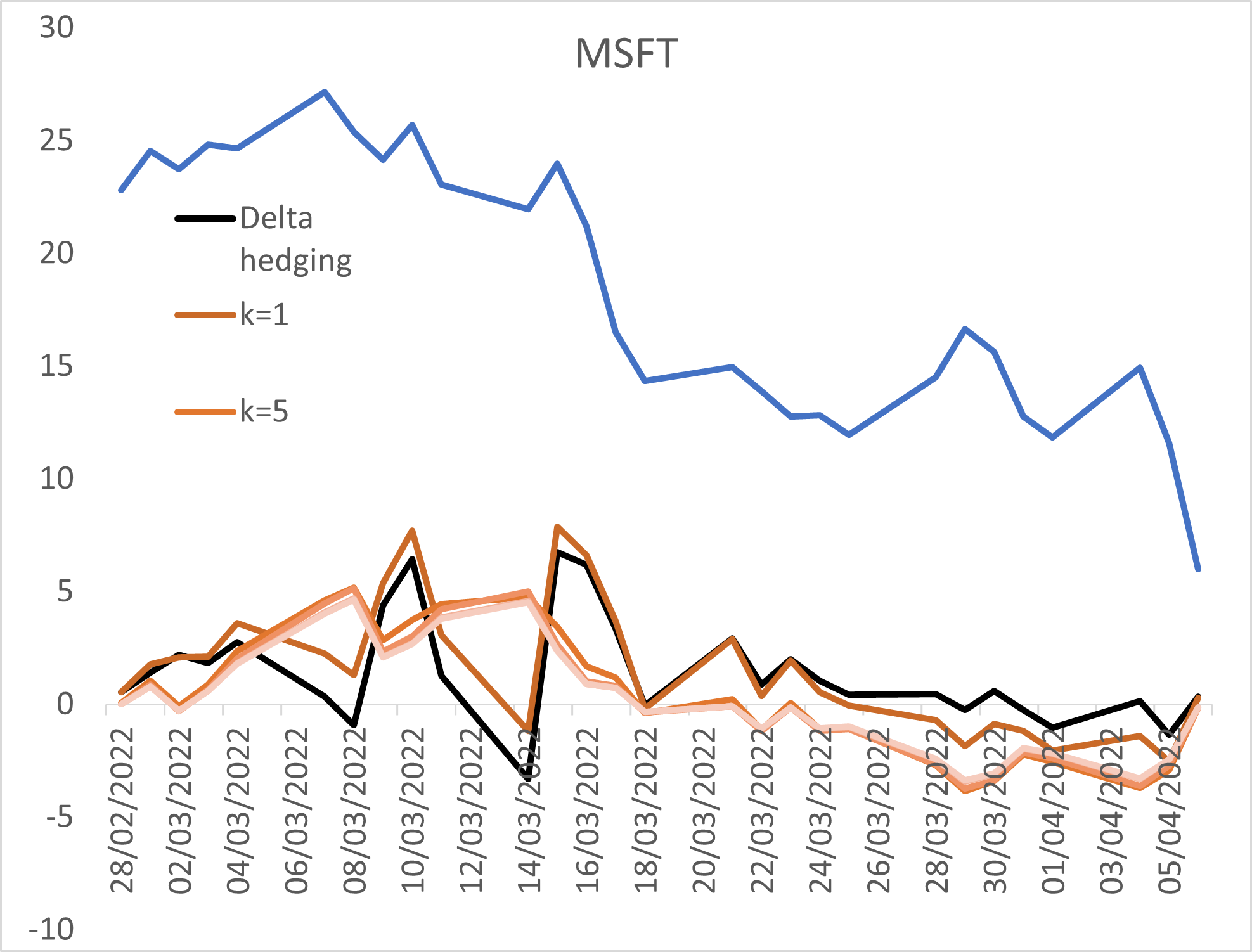}
\includegraphics[width=4cm]{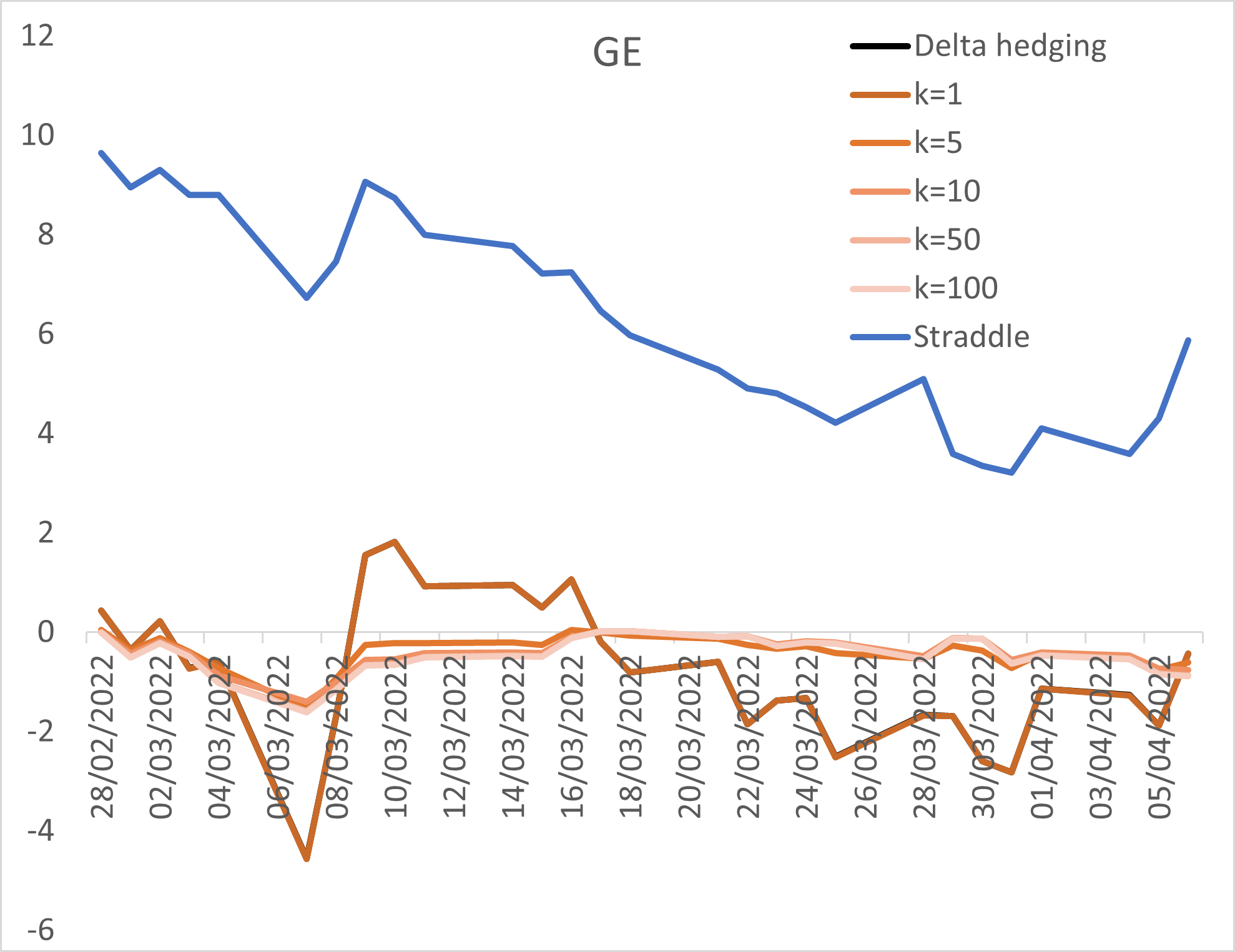}
\includegraphics[width=4cm]{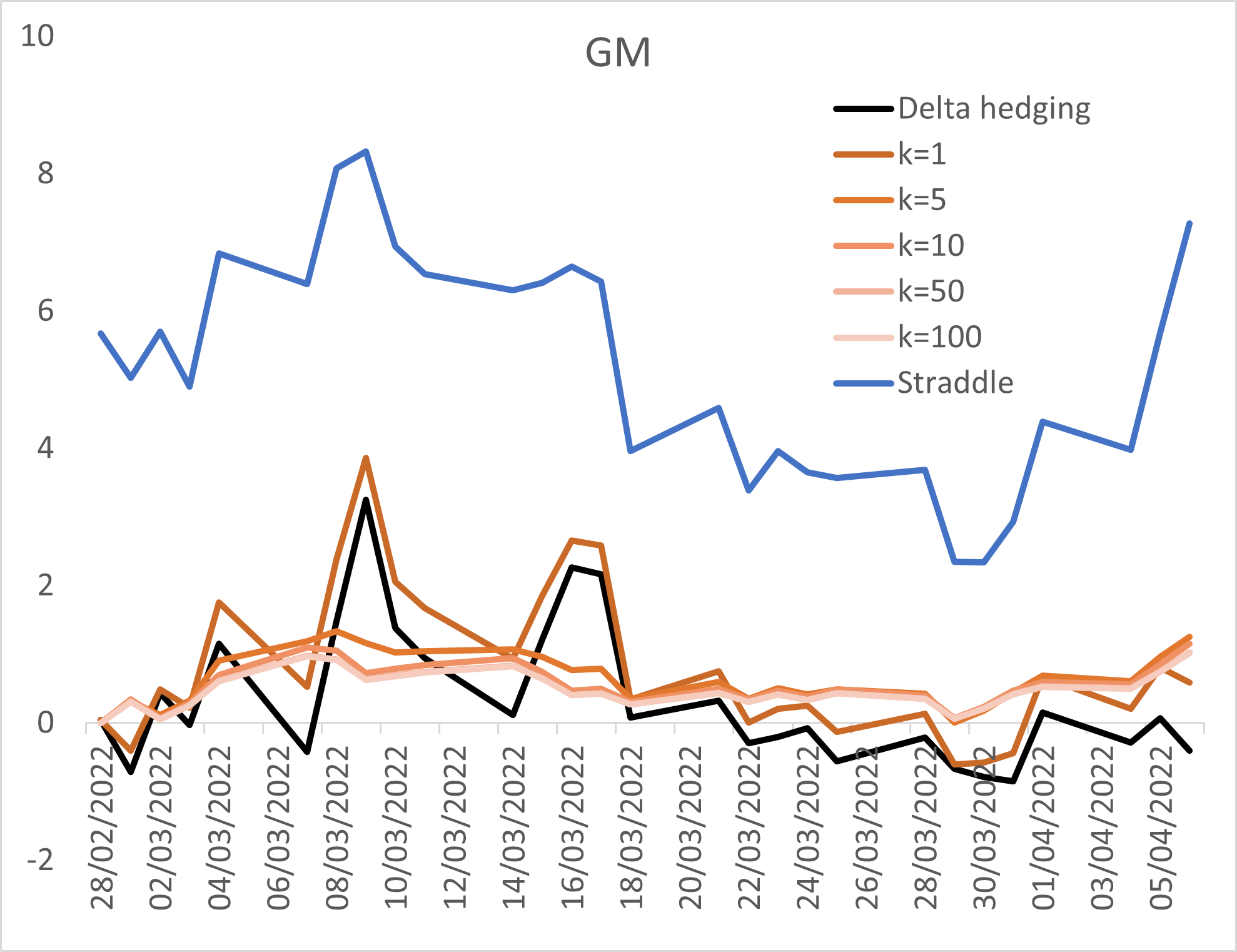}
\includegraphics[width=4cm]{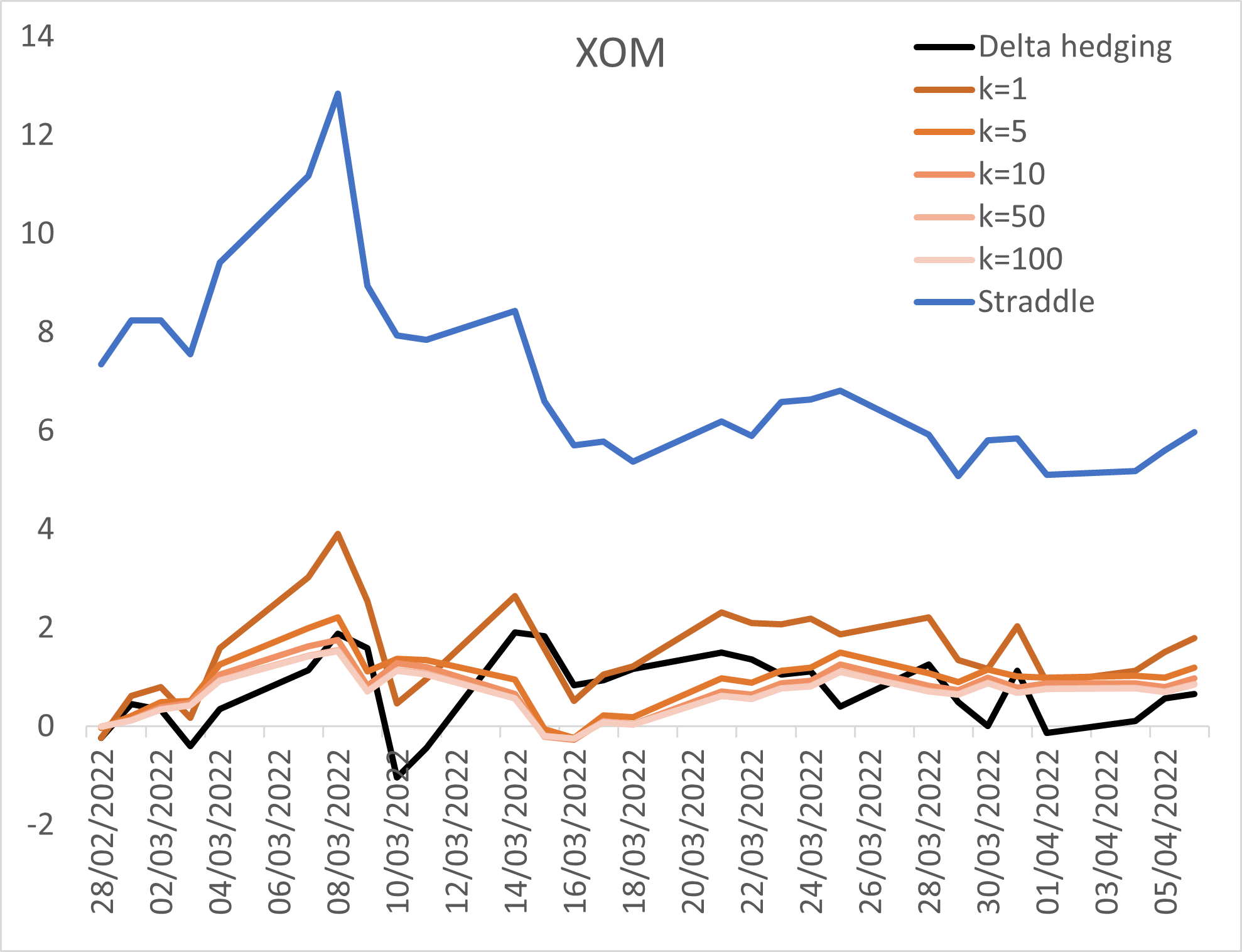}
\caption{Option hedging PnL for $k=1,2,10, 50, 100$, for AAPL, MSFT, GE, GM and XOM with risk tolerance $\lambda=5\%$.}
\label{fig:options}
\end{figure}

The resulting hedge weights for zero option delta are shown in Figure \ref{fig:optionweights}, and the PnL trajectories are shown in Figure \ref{fig:options}. The modified hedge  (\ref{optionweight}) generally produced lower PnL volatility than pure delta hedging. The typical picture is that the modified hedge performs very similarly to the delta hedge when $k=1$, and the PnL volatility then decreases with increasing $k$. For increasing values of the penalty exponent $k$, the modified hedging PnLs showed marked convergence. Higher exponents $k=50$ and $k=100$ were  remarkably close in all examples, indicating that the effect of increasing $k$ levels off reasonably quickly. This is true both on the level of weights, as seen in Figure \ref{fig:optionweights}, and PnL trajectories, as seen in Figure \ref{fig:options}.

The resulting hedging PnL is noticably higher where the guess of the drift term was correct, namely XOM. It was similarly higher for GM, where the guess was incorrect. For the remaining symbols, the final PnL was markedly close for all trajectories.

\section{Conclusions}

The methods presented here are intended to shed new light on optimal portfolio construction in the presence of fat tails, especially in the context of liability-driven portfolios. The sort of portfolios considered include classical LDI pension fund and insurance portfolios, with a stream of essentially fixed liabilities, but also the sort of portfolios that arise in client-facing derivatives books, where the liabilities comprise of delta one products or options written to clients.

The risk measures employed are a continuous family parametrized by a single parameter $k$, and they range from variance, when $k=1$, through maximum-drawdown-like measures as $k\rightarrow \infty$. These risk measures were studied extensively in \cite{PLP}, but not in the context of liability-driven portfolios.

The limit of $k \rightarrow \infty$ results in a new risk measure, XD. XD is significantly more sensitive to extreme returns than CVaR, and yet it can be estimated robustly as the limit of high order moments. Our optimal portfolios in the $k \rightarrow \infty$ limit optimize the return per unit of XD.

The general pattern, which holds across all the examples we show, is summarized in equation (\ref{weight}). Briefly, the weight of an orthogonal component is comprised of a risk term, proportional to the ratio of the return to the relevant moment of the component returns and the risk appetite parameter, and a hedging term neutralizing as much of the liability as possible, all raised to the power of $1/(2k-1)$.

The limit of $k\rightarrow \infty$  reveals some non-trivial results, namely that the liability term reduces the hurdle rate for the component return, and that the resulting effective hurdle rate tends exponentially to $-\infty$ for $k \rightarrow \infty$.

There are interesting implications of this sort of portfolio construction to Constant Proportion Portfolio Insurance (CPPI) products \cite{cppi1}. The main risk in CPPI products, gap risk, is due to unexpected large jumps either in the portfolio, or in its benchmark; with our portfolio construction, we aim to manage both tail risk in the underlying portfolio and the tail risk in the benchmark. Hence there are interesting applications to be explored in that direction. 

Along those lines, it is perhaps not surprising that the value of the CPPI cushion also follow a power law \cite{cppi2, cppi3}. The power law and the resulting analysis are nonetheless somewhat different, due to the asymmetry of the CPPI payout where the hedging term decays for sufficiently outperforming portfolio components, which is not the case in our formulation.

The actual choice of $k$ in a practical situation is subject to multiple trade-offs. The large $k$ limit is of obvious interest, due to its link to the maximum drawdown. The flip side of that is, however, difficult to estimate any statistics for. Leaving aside whether it is even theoretically possible to predict future extremals of an unknown distribution from its history, we saw in practical examples that the convergence is reasonably rapid. In all examples, there is a rapid change in the optimal portfolio as $k$ increases from $k=1$ to $k \sim 10$, follow by convergence by $k \sim 100$. We can therefore reasonably conclude that the limit, insofar as it can be estimated from the historical distribution, converges rapidly.

On the other hand, the theory behind the $k\rightarrow \infty$ limit can be summarised in five words - \emph{"to manage drawdowns, diversify everything"}, even if we can not accurately estimate moments and extremals.

\end{document}